# 2D Magnetic Semiconductors via Substitutional Doping of Transition Metal Dichalcogenides

**Mengqi Fang1 and Eui-Hyeok Yang1,2,***

1. Department of Mechanical Engineering, Stevens Institute of Technology, Hoboken, New Jersey 07030, United States
2. Center for Quantum Science and Engineering, Stevens Institute of Technology, Hoboken, New Jersey 07030, United States
* Address correspondence to: eyang@stevens.edu

**Abstract:** Transition metal dichalcogenides (TMDs) are two-dimensional (2D) materials with remarkable electrical, optical and chemical properties. One promising strategy to tailor TMD properties of TMDs is to create alloys through dopant-induced modification. Dopants can introduce additional states within the bandgap of TMDs, leading to changes in their optical, electronic, and magnetic properties. This paper overviews chemical vapor deposition (CVD) methods to introduce dopants into TMD monolayers. The advantages and limitations and their impacts on the doped TMDs' structural, electrical, optical, and magnetic properties are discussed. The dopants in TMDs modify the density and type of carriers in the material, thereby influencing the optical properties of the materials. The TMDs' magnetic moment and circular dichroism are also strongly affected by doping, which enhances the magnetic signal in the material. Finally, we highlight the different doping-induced magnetic properties of TMDs, including superexchange-induced ferromagnetism and valley Zeeman shift. Overall, this review paper provides a comprehensive summary of magnetic TMDs synthesized via CVD, which can guide future research on doped TMDs for various applications, such as spintronics, optoelectronics, and magnetic memory devices.

## 1. Introduction

The van der Waals crystals are two-dimensional (2D) material systems with atomically thin, layered structures. Recent discoveries in 2D materials include ferromagnetism in atomically thin layers of chromium-based alloys [1]. For example, $Cr_2Ge_2Te_6$ demonstrates a ferromagnetic order that remarkably influences the transition temperature, even with small magnetic fields [2]. Mechanically exfoliated monolayer chromium triiodide ($CrI_3$) also achieves ferromagnet properties with out-of-plane spin orientation under 45 K [3]. While these van der Waals ferromagnets remain either metallic or insulating, spintronics and solid-state quantum information science applications benefit from semiconductors with ferromagnetic properties known as dilute magnetic semiconductors (DMS). These DMS materials have been extensively researched for decades in their bulk form, of which typical representatives are (Cd, Mn)Te, (Zn, Co)O, and (Pb, Eu)S [4]. Several theoretical studies have predicted that DMS based on TMDs would exhibit ferromagnetic behaviors even at room temperature, which is a fundamental requirement for practical applications [5–10]. For example, first-principles calculations anticipated that Fe and V-doped $MoSe_2$ would exhibit room-temperature out-of-plane ferromagnets at high atomic substitution [5].

Several experimental studies have utilized ex-situ techniques to dope transition metal atoms (V, Cr, Fe, etc.) in post-processing techniques into TMDs, thereby creating the first evidence that 2D DMS can be realized [11], while resulting either in a Curie temperature well below room temperature or in random local clustering of magnetic precipitations. Other attempts to incorporate magnetic atoms directly in situ during growth relied upon converting bulk TMD crystals into DMS. For example, Fe was doped in $SnS_2$ bulk crystal via a direct vapor-phase method, while mechanical exfoliation was required to reveal van der Waals DMS [12]. Recently, direct in-situ growth of transition metal-doped TMDs via chemical vapor deposition (CVD) overcame some of the difficulties of post-processing conversion approaches [13]. For example, rhenium (Re), vanadium (V), or iron (Fe) was proven to substitute either molybdenum (Mo), tungsten (W) sites in monolayer $MoS_2$, $WS_2$, and $WSe_2$ crystalline [14].

Combining magnetism with semiconducting properties, 2D magnetic DMS offers intriguing applications, including magneto-optoelectronic devices for spintronics, optics, and the quantum field. Fe-doped $SnS_2$ homojunction device was reported with an unsaturated magnetoresistance (MR) of 1800% [15]. Giant magnetic circular dichroism was demonstrated by the magneto-optical effects in monolayer, bilayer, and trilayer DMS, leading to significant MO Kerr rotation and Faraday rotation angles. [16]. Reports on device applications using 2D magnets include the magnetic tunnel junction in heterostructure devices with 2D layers of $Fe_3GeTe_2$ [17], $Cr_2Ge_2Te_6$ [18], and $CrI_3$ [19]. Ferromagnetic properties in doped TMDs provide potential opportunities for many groundbreaking applications. Here, we review the synthesis, characterization, magnetic properties, and applications of 2D magnetic semiconductors.

## 2. Ferromagnetism in 2D limits

Ferromagnetism in 2D limits can be induced by substituting the atoms in the d-block of the periodic table with the original transition metal atoms or chalcogen atoms, realizing 2D DMS materials. Transition metal dopants and transition atoms in TMD lattices are polarized primarily by their localized 3d and 5d electrons. For instance, based on the density functional theory, a hybridization of localized transition metal-3d, delocalized Se-4p, or W-5d states results in electronic structures of spin-up and spin-down channels in the $WSe_2$ lattice, leading to the ferromagnetic coupling between the doped transition metal atom spins and Se or W spins [7]. Similarly, due to low formation energy, Mn dopants in $MoS_2$ tend to replace Mo atoms in Mn-doped $MoS_2$ crystals [20]. Upon substitution, a portion of the Mn dopants' valence electrons ($3d^64s^1$) move to the nearest S atoms, and the remaining valence electrons cause spin polarization, creating a magnetic moment [21]. Dopant concentrations also affect the magnetic-exchange coupling related to ferromagnetic and antiferromagnetic properties. For example, when two Mn dopants are close to each other, the ferromagnetic state is preferred. However, the Mn dopants exhibit strong spin polarization and antiferromagnetic coupling with the S atoms when their separation distance is less than 16.53 Å, resulting in the same spin-polarization direction [21,22]. Besides, $MoSe_2$ is doped with Ni, V, Mn, Cr, Fe, Co, Ni and two dopant atoms (Ti, V, Mn, Cr, Fe, Co, Ni) in another article, resulting in a weakly magnetic/non-magnetic (NM) state in $MoSe_2$ with Ni and Ti as dopants. V, Cr, Mn, Fe, and Co have magnetic moments exceeding 1.6 μB per dopant [5].The electronic band structures also significantly change in d-orbital hybridization after Fe is substitutionally doped in monolayer $MoS_2$ with a further enhanced valley Zeeman splitting [23]. The effect of hybridization between the conduction band and d band in the transition metal dopants and the TMD lattice leads to ferromagnetism and antiferromagnetism in 2D doping TMDs.

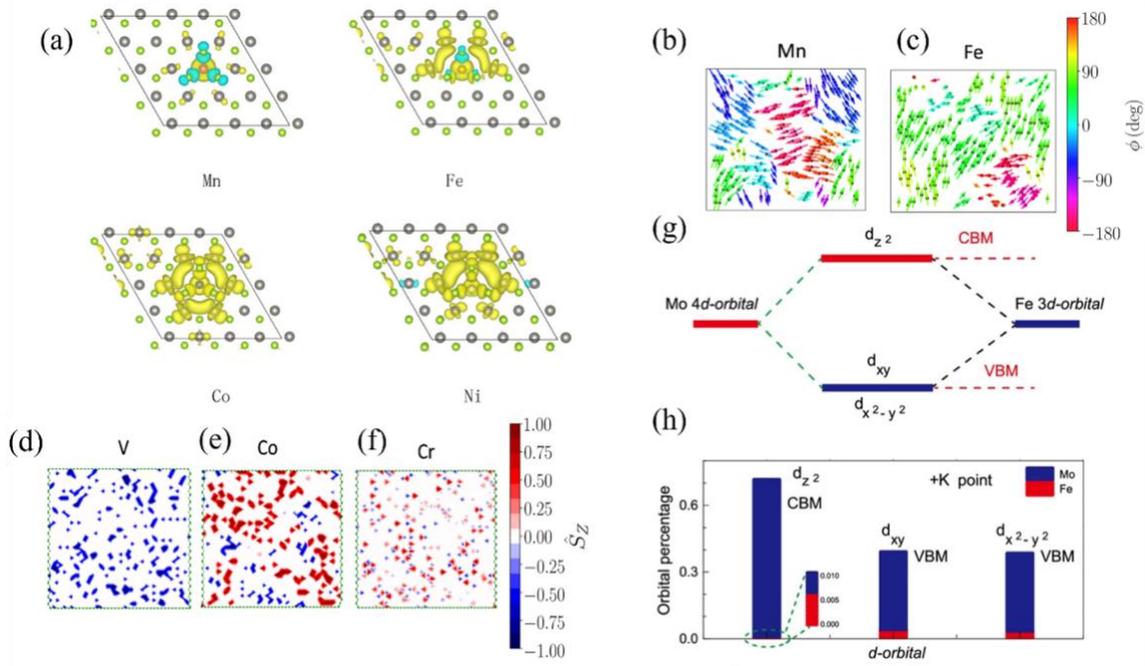

**Figure 1.** (a) Spin density for a single transition metal dopant atom (Mn, Fe, Co, Ni) in a monolayer $WSe_2$ supercell with a 6.25% doping concentration of transition metals. Yellow and cyan isosurfaces represent positive and negative spin densities [7]. (b-f) The magnetically ordered state of a Mn (b), Fe (c), V (d), Co (e), and Cr-doped (f) sample of $MoSe_2$ at an atomic substitution of 15%, at a temperature of 5 K [5]. (g) Diagram of Fe and Mo atoms hybridized in d-orbitals for a 5x5 $MoS_2$ supercell with one Fe atom instead of one Mo atom. (h) Percentage of hybridized d-orbitals of Fe-doped monolayer $MoS_2$ for each isolator d-orbital. The valence band maximum is predominantly composed of the $d_{x^2-y^2}$ and $d_{xy}$ orbitals of both Mo and Fe, whereas the conduction band minimum is primarily dominated by the $d_{z^2}$ orbitals. Although the Fe d-orbitals contribute to the orbital hybridization, their relative fractions are comparatively small compared to Mo d-orbitals [23]

### 3. Synthesis

Several synthesis strategies have been demonstrated, including CVD, physical vapor deposition (PVD), atomic layer deposition (ALD), molecular beam epitaxy (MBE), and pulsed laser deposition (PLD) [24]. This review focuses on the synthesis of 2D DMS via CVD. Due to the relative simplicity of CVD systems in the laboratory and the ability to deliver large single-crystal domains, CVD systems are commonly used to synthesize 2D TMDs [25]. Typically, the transition metal source sits in the middle of the heating zone, and the chalcogen source is placed upstream of the carrier gas (typically Ar or $N_2$, usually combined with $H_2$). Heating the furnace to 700-1000 °C causes the vaporization of transition metal and chalcogen sources, leading to the formation of nanoparticles diffusing in the carrying gas, which then condenses on the substrate. Doping can be achieved by diffusing dopants into TMDs during the vaporization of precursor, making CVD-growth an effective means to generate TMD alloys [26,27].

*3.1. LPCVD growth and doping*

3.1.1. Solid source-based growth and doping of TMDs

The solid source-based TMD growth requires a mixture of solid dopant powder, such as elemental substance, metal salt, and metal oxide, with a transition metal source. This growth is realized based on the simple mixing of precursors, which does not involve additional multiplex pre-operating procedures or complex pressure and flow control. For example, Fe-doped $MoS_2$ can be synthesized using the $MoO_3$ and $FeS_2$ powders with different Fe-to-Mo molar ratios of 0.12, 0.24, 0.48, and 0.4 [23]. Cr-doped $WTe_2$ crystals were grown via a two-step Te flux. The W and Cr powders and Te granules were mixed and heated at 1050 °C for 24 hours, then Cr atoms were diffused into layered semimetal Td-WTe$_2$ [28]. Mn atoms were successfully in situ doped into the MoS$_2$ lattice by introducing the Mn$_2$(CO)$_{10}$ precursor upstream, whereby the incorporation of Mn in MoS$_2$ was enabled by inert substrates but not reactive surface terminations, which led to the formation of defective MoS$_2$. [13]. Another research used Fe$_3$O$_4$ particles cast on the substrate as the dopant source, permitting substitutional doping of Fe atoms into Mo or W sites in TMD crystals [29,30].

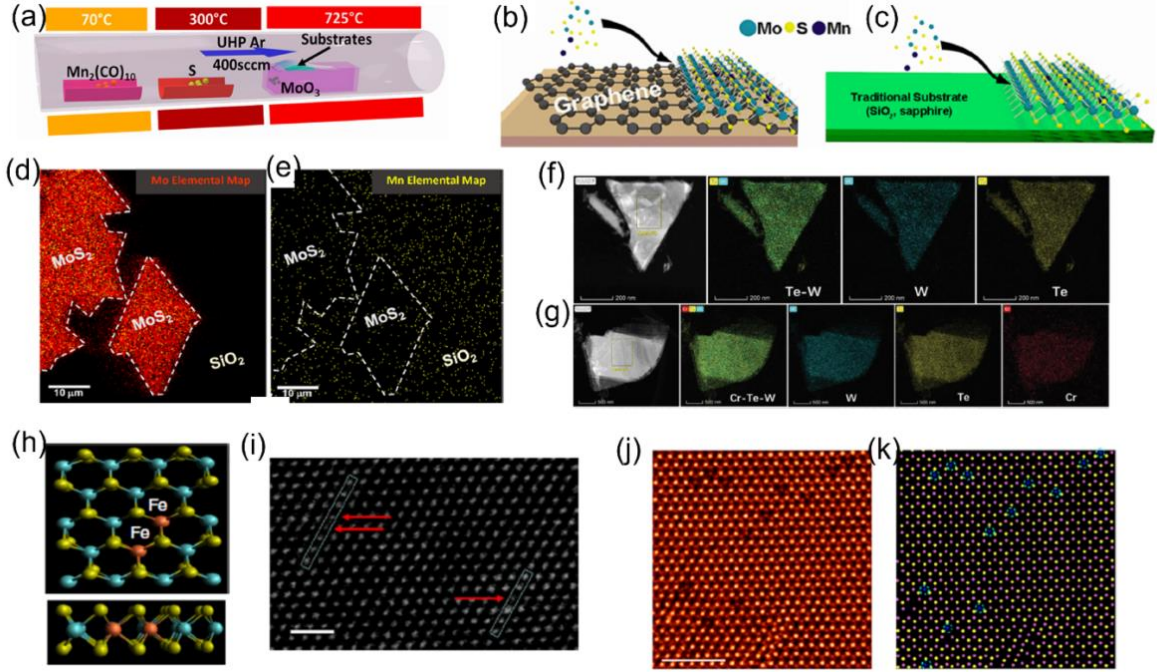

**Figure 2.** (a) Schematic of powder vaporization techniques used to synthesize manganese-doped and undoped MoS$_2$ [13]. Schematic of growth substrates, including (b) Graphene and (c) Insulating substrate (sapphire and SiO$_2$). (d) Element analysis of Mn-doped MoS$_2$ with (e) Mn concentrations (yellow pixels) are equal (or higher) to those found in the MoS$_2$ areas, which suggests that Mn may be bound to the substrate surface rather than incorporated into it [13]. (f)(g) STEM-HAADF images of (f) Pure WTe$_2$ and (g) Cr$_{0.02}$–WTe$_2$ samples with their corresponding element mapping images [28]. (h) Schematics of the monolayer Fe:MoS$_2$, with S, Mo, and Fe atoms denoted by green, blue, and red spheres. (i) Contrast-corrected STEM image of monolayer Fe:MoS$_2$ [31]. (j) Typical atomic-resolution STEM-ADF image of Fe-doped 1H-MoS$_2$ monolayer. (k) Corresponding atomic mode showing the distribution of Fe dopant atoms in (a). Blue, pink, and yellow spheres represent Fe, Mo, and S atoms [23].

3.1.2. Liquid-assisted growth and doping of TMDs

The liquid-assisted growth uses a metal salt aqueous solution or metal suspension directly spun on the substrate. Although it requires more steps than using the solid source, it enables a more homogeneous mixing of precursors, contributing to more uniform doping in the growing TMD lattice than the solid source-based doping. For example, v-doped monolayer WSe$_2$ demonstrated ferromagnetic domains by controlling the atomic ratio of V to W in precursor solution from 0.1% to 40%. Furthermore, in this growth, the doping concentration was controlled from 1% to a relatively high atom weight percent [32]. Lastly, the one-pot mixed-salt-intermediated CVD method was employed to synthesize single-crystal magnetic group VIII transition metal-doped MoSe$_2$, which demonstrated excellent controllability and reproducibility, with Fe-doped MoSe$_2$ monolayers obtained by adjusting the FeCl$_3$/Na$_2$MoO$_4$ ratios to control Fe-doping concentrations ranging from 0.93% to 6.10% [33].

*3.2. MOCVD growth and doping*

The metal-organic chemical vapor deposition (MOCVD) uses a pulsed precursor gas vapor source. It is a highly complex process for growing wafer-scale 2D crystalline layers with an excellent uniformity of deposition rates, dopant concentrations, and layer thickness [34], which is crucial to for practical nanoelectronics and optoelectronics applications [35]. For example, large-area Nb-doped monolayer $MoS_2$ was grown by employing molybdenum hexacarbonyl ($Mo(CO)_6$) and diethyl sulfide (DES) as Mo and S precursors [35]. Multilayer V-doped $WSe_2$ has been demonstrated on quasi-freestanding epitaxial graphene on 6H-SiC [36]. $WSe_2$ films with coalesced 2D surfaces were doped with 0.5-1.1 at % Re atoms in situ in another systematic study: a precisely controlled doping concentration was achieved by tuning the precursor partial pressure, with decreased domain size with increased doping concentration [37].

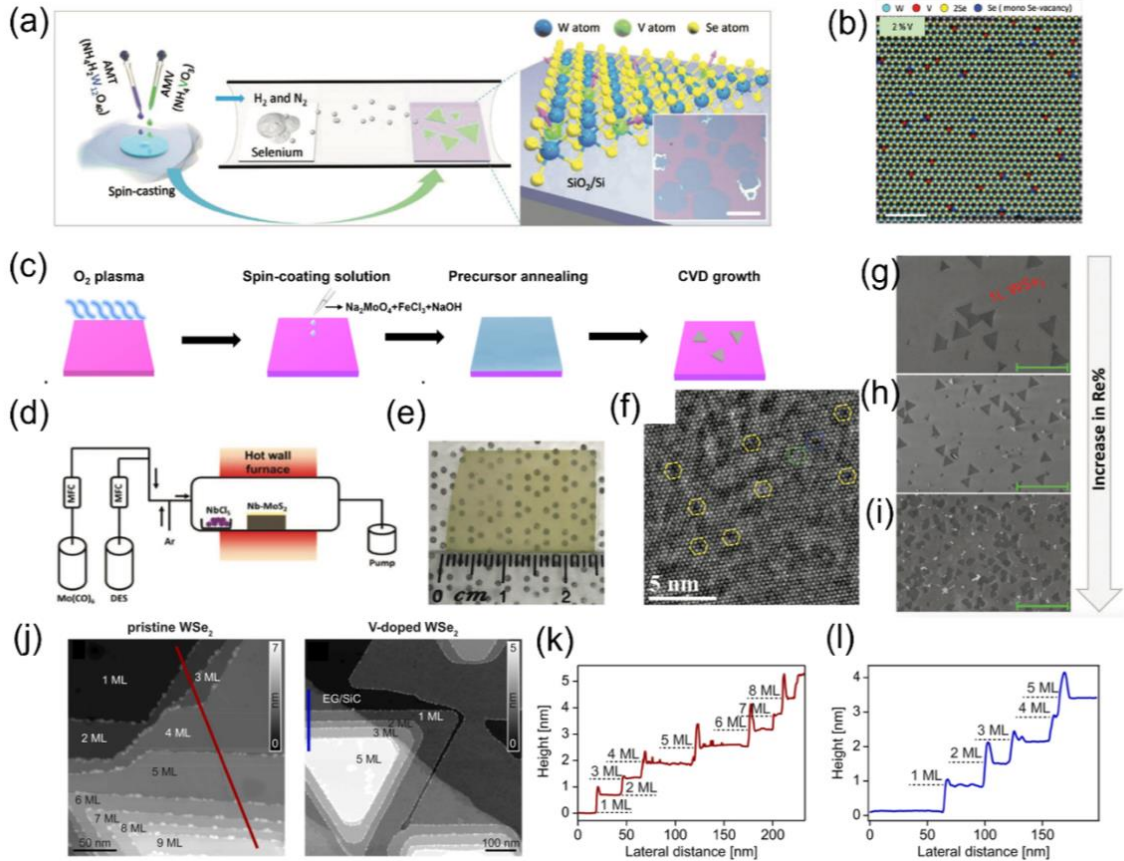

**Figure 3.** (a) Schematic of the V-doped $WSe_2$ synthesis using liquid precursor mixing W with V. (b) False-color Wiener-filtered STEM images of 2% V-doped $WSe_2$ [32]. (c) Process sequence of Fe-doped $MoSe_2$ using a one-pot mixed-salt-intermediated CVD method [33]. (d) Schematic of the MOCVD setup. (e) Optical image of uniformly grown Nb-doped $MoS_2$, the scale bar is 1 cm [35]. (f) HRTEM image shows a periodic atomic arrangement where Re atoms are introduced in the $WSe_2$ lattice. (g) SEM images of Re:$WSe_2$ monolayers with 0.5% (h) and 1.1% (i) doping concentrations. Re-doped $WSe_2$ films reveal significant changes in the surface morphology following the introduction of Re atoms [37]. (j) STM image (VB = 1.6 V) of undoped and V-doped multilayer WSe2, with the number of $WSe_2$ layers labeled, providing valuable insights into the electronic and structural properties of $WSe_2$ materials. (k)(l) Height profiles along the lines shown in the corresponding SEM images of undoped (k) and V-doped $WSe_2$ (l) [36].

As summarized in Table 1, while solid source-based CVD growth and doping are convenient means to grow various TMDs, it requires vaporizing the metal-oxide precursors. The liquid phase-assisted CVD growth and doping can achieve a more homogeneous mixing of dopants and transition metal sources at the molecular level than the solid source-based CVD growth. MOCVD provides precise control of precursor supply to grow TMDs on a wafer scale. The metal-organic precursors can easily be evaporated to the vapor phase [38], permitting a lower growth temperature than LPCVD growth. However, residual carbon contamination may occur with some precursors in the grown 2D film [39].

**Table 1.** Comparison of growth methods.

| Material type | Dopant | Synthesis methods | Size | Thickness | Doping concentrations | Semiconductor type after doping | Ref |
|---|---|---|---|---|---|---|---|
| MoS$_2$ | Co,Cr | Solid sourse CVD | ~20um | monolayer | Co 1% Cr 0.3% | p-type | [40] |
| MoS$_2$ | Fe | Solid source CVD | ~20um | monolayer | 0.3~0.5% | — | [29] |
| MoS$_2$ | Fe | Solid source CVD | ~30um | monolayer | 0.40% | — | [23] |
| WSe$_2$ | V | Liquid-phase assistant CVD | >50 um | monolayer | 0.5%-10% | p-type | [32] |
| Td-WTe$_2$ | Cr | two-step Te flux CVD | ~1um | bulk | 2% | — | [28] |
| MoS$_2$ | Mn | Solid source CVD | ~200nm | monolayer | 2% | — | [13] |
| WS$_2$ | V | Liquid-phase assistant CVD | — | monolayer | 0.4-12% | p-type | [41] |
| MoSe$_2$ | Fe | Liquid-phase assistant CVD | ~40 um | monolayer | 0.93-6.1% | n-type | [33] |
| MoTe$_2$ | Cr | Solid source CVD | >1um | 2H bulk | 2.1-4.3% | p-type | [42] |
| MoS$_2$ | Re | Solid source CVD | ~15um | monolayer | 1% | n-type | [14] |
| MoS$_2$ | Nb | MOCVD | Wafer-scale | monolayer | 5% | — | [35] |
| WSe$_2$ | V | MOCVD | — | monolayer to multilayer | 0.44% | p-type | [36] |
| WSe$_2$ | Re | MOCVD | 450-500 nm | Monolayer | 0.5%-1.1% | n-type | [37] |

## 4. Characterization

*4.1. Optical properties*

The electron-phonon coupling strengths are impacted by the modification of the A'1 vibrational mode (out-of-plane vibration of chalcogen atoms), which is caused by the electron density shift resulting from the substitutional doping [43–45]. For example, Fe-doped MoS$_2$ showed the n-type doping effect, which is revealed by a blue shift of the A$_{1g}$ peak compared to undoped MoS$_2$ [30]. Co or Cr-doped MoS$_2$ showed a moderate blue shift of the A$_{1g}$ peak, while unaffecting the E$^1_{2g}$ peak (in-plane vibration mode of chalcogen atoms), showing the sensitivity of A$_{1g}$ peak to dopant atoms [40]. The density and type of carriers can be modified after doping, affecting the PL spectra of the materials dramatically. For example, the Re-doping of WSe$_2$ showed a PL redshift by 40 meV and quenching by a factor of two, caused by non-radiative trion recombination [37]. A similar redshift was observed in Nb-doped MoS$_2$ with a ~150% intensity enhancement due to impurity state formation caused by the Nb atoms [35]. On the other hand, Fe-doped WS$_2$ showed a PL blueshift by 13 ± 0.5 meV and PL quenching by 40%. Fe-doped MoS$_2$ exhibited a PL quenching of 35% and a redshift of 29 ± 0.5 meV [30]. The difference between Fe:MoS$_2$ and Fe:WS$_2$ is attributed to the interaction of neutral excitons and negative/positive trions [46].

*4.2. Magnetic properties*

In TMD crystals, direct interband transitions at the K and K' valleys are coupled exclusively to the left and right circularly polarized light [47]. Furthermore, the inversion symmetry enhanced by dopant atoms leads to the existence of a magnetic moment within the orbit of the electrons. Therefore, the valley magnetic moment can be controlled in monolayer TMDs by circularly polarized light and the out-of-plane magnetic field, permitting control of the valley degree of freedom [48]. Strong circular dichroism (CD) was observed in the Fe-related peak at 2.28eV from monolayer Fe-doped MoS$_2$ when opposite circularly polarized lights were applied. The magnetic circular dichroism (MCD), as a function of increasing and decreasing magnetic fields, also showed a distinct hysteresis loop [29]. Another report showed a valley Zeeman shift of Fe-doped MoS$_2$. At zero field, the PL spectra were completely overlapped when σ− and σ+ light were applied; however, they split at high magnetic fields with inverse shifting directions for opposite magnetic field directions [23].

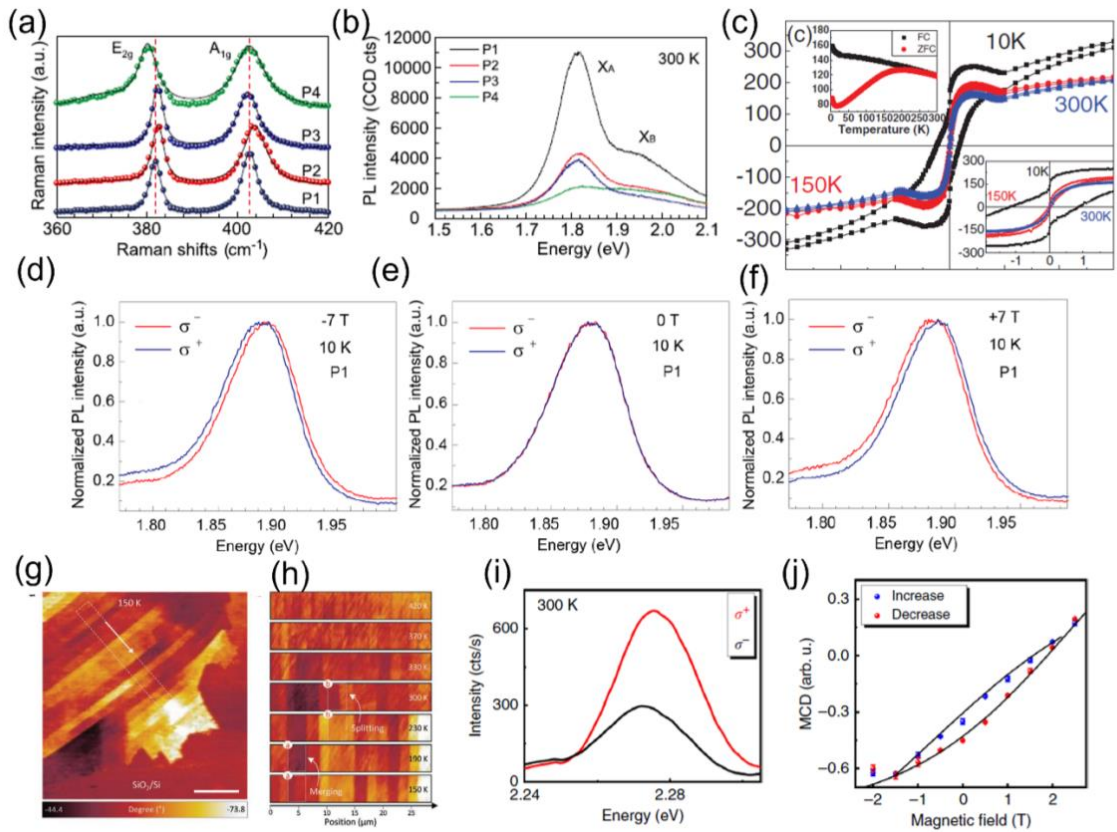

**Figure 4.** (a) Raman and (b) PL spectra of Fe-doped monolayer MoS$_2$ at room temperature [23]. (c) The hysteresis loop of 1 at% Nb-4 at% Co codoped WSe$_2$, along with zero-field-cooled (ZFC) and field-cooled (FC) curves shown in the top left inset, and magnified M-H curves in the bottom right inset, provide a characterization of the magnetic properties at different temperatures [49]. (d) The normalized raw polarization-resolved valley-exciton photoluminescence (PL) spectra of Fe-doped monolayer MoS$_2$ without a magnetic field (e) and under ±7 T (d-f) at 10 K [23]. (g) MFM phase image of 0.1% V-doped WSe$_2$ taken at 150 K. (h) Temperature-dependent transition of magnetic

domains [32]. (i) The red and black spectra represent the Fe-related spontaneous emissions under excitation with opposite circularly polarized light states at 300 K. (j) Corresponding magnetic circular dichroism (MCD) as a function of increasing (blue circles) and decreasing (red circles) magnetic fields, recorded at 4 K [31].

Goodenough-Kanamori-Anderson rules state that superexchange can lead to ferromagnetic short-range coupling between overlapping orbitals [50]. In Nb and Co codoped $WSe_2$, the magnetic signal was strongly enhanced with the doping of 4 at% Co. The saturation magnetization (Ms) reached 63.6 emu cm$^{−3}$ at 10 K [49]. Magnetization versus magnetic field strength (M−H) test for Cr-doped $MoTe_2$ showed a significant enhancement in an out-of-plane magnetic field with Curie temperature around 275 K. The maximum Ms value occurred at 4.78 emu g$^{−1}$ for the bulk Cr-4.3% doped $MoTe_2$ [42]. Semiconducting V-doped $WSe_2$ monolayers showed temperature-dependent ferromagnetic domains with long-range order. Magnetic force microscopy (MFM) was used to detect the ferromagnetic domain stripes with a solid contrast, where the splitting and merging phenomenon of domains was observed with the temperature changes from low to high temperature. The magnetic domain walls retained the vestige even at 420 K [51]. The reported magnetization and coercivity in doped TMDs are summarized in Table 2.

Table 2. The magnetization and coercivity in different transition metal-doped TMDs.

| Material | Dopant concentration | Saturation magnetization | Coercivity | Temperature | Reference |
|---|---|---|---|---|---|
| Fe doped $SnS_2$ | 2.10% | 3.49*10-3 emug$^{-1}$ | 400 Oe | 300K | [12] |
| Co-Cr doped $MoS_2$ | 1% Co, 0.3% Cr | 0.4 emu cm$^{−3}$ | 100 Oe | 300K | [40] |
| V doped $MoTe_2$ | 0.30% | 0.6 μemu cm$^{−2}$ | - | 300K | [52] |
| Co and Nb doped $WSe_2$ | 4% | 60.62 emu g$^{−1}$ | 1.2 kOe | 10K | [49] |
| Cr doped Td-$WTe_2$ | 1% | 4.20 emu g$^{−1}$ | - | 3K | [28] |
| V doped $WS_2$ | 2% | 2.85*10-5 emu cm$^{-2}$ | 180 Oe | 50K | [41] |
| Mn doped $MoSe_2$ | 6.10% | 2*10-5 emu g$^{-1}$ | - | 300K | [33] |
| Cr doped 2H-$MoTe_2$ | 2.50% | 4.78 emu g$^{−1}$ | 6322 Oe | 3K | [42] |
| V doped $MoS_2$ | 5% | 0.067 emu g$^{−1}$ | 1870 Oe | 10K | [53] |

## 5. Conclusions

Doping of TMDs is a popular topic since the external atoms can bring or enhance multiple properties of origin TMDs. In magnetic doping, the promoted magnetic properties due to the large atomic mass of doped transition metals can shed new light on their potential applications. This review has introduced different doping methods via CVD, including solid source-based doping, liquid-assisted doping, and

MOCVD methods. Dopant-induced optical and magnetic properties of doped TMDs were also summarized.

The exploration of magnetically doped TMDs poses several challenges. First, the uniform and large-area TMD synthesis remains challenging due to the difficulty in controlling the flow of vaporized precursors and uncontrollable domain size [54]. Though room temperature ferromagnets were demonstrated, most of the magnetic TMDs show low curie temperatures hampering practical applications of these materials. Optimizing the exchange interaction and magnetic anisotropy would be key parameters to enhance the critical temperatures [1]. Another challenge includes randomly distributed dopants in the crystal lattice, causing localized magnetic fields. The long-range behavior of magnets is contributed to the finite lattice size, concentrated and ordered positions of magnetic ions, and different saturation magnetizations caused by different kinds of dopants. [5]. The topological boundary states, magnetic proximity effects and van der Waals heterostructures are emerging fields utilizing magnetic 2D TMDs. With groundbreaking research in the coming years, transition metal-doped TMDs are promising materials in optomagnetism, spintronic devices, and quantum information science.

This section is not mandatory but can be added to the manuscript if the discussion is unusually long or complex.

**References**

1. Burch, K.S.; Mandrus, D.; Park, J.-G. Magnetism in two-dimensional van der Waals materials. *Nature* **2018**, *563*, 47–52.
2. Gong, C.; Li, L.; Li, Z.; Ji, H.; Stern, A.; Xia, Y.; Cao, T.; Bao, W.; Wang, C.; Wang, Y.; et al. Discovery of intrinsic ferromagnetism in two-dimensional van der Waals crystals. *Nature* **2017**, *546*, 265–269, doi:10.1038/nature22060.
3. Huang, B.; Clark, G.; Navarro-Moratalla, E.; Klein, D.R.; Cheng, R.; Seyler, K.L.; Zhong, D.; Schmidgall, E.; McGuire, M.A.; Cobden, D.H.; et al. Layer-dependent ferromagnetism in a van der Waals crystal down to the monolayer limit. *Nature* **2017**, *546*, 270–273, doi:10.1038/nature22391.
4. Dietl, T.; Bonanni, A.; Ohno, H. Families of magnetic semiconductors - An overview. *J. Semicond.* **2019**, *40*, doi:10.1088/1674-4926/40/8/080301.
5. Tiwari, S.; de Put, M.L.; Sorée, B.; Vandenberghe, W.G. Magnetic order and critical temperature of substitutionally doped transition metal dichalcogenide monolayers. *npj 2D Mater. Appl.* **2021**, *5*, 1–7.
6. Fan, X.-L.; An, Y.-R.; Guo, W.-J. Ferromagnetism in transitional metal-doped $MoS_2$ monolayer. *Nanoscale Res. Lett.* **2016**, *11*, 1–10.
7. Zhao, X.; Chen, P.; Wang, T. Controlled electronic and magnetic properties of $WSe_2$ monolayers by doping transition-metal atoms. *Superlattices Microstruct.* **2016**, *100*, 252–257.
8. Cheng, Y.C.; Zhu, Z.Y.; Mi, W.B.; Guo, Z.B.; Schwingenschlögl, U. Prediction of two-dimensional diluted magnetic semiconductors: Doped monolayer $MoS_2$ systems. *Phys. Rev. B - Condens. Matter Mater. Phys.* **2013**, *87*, 2–5, doi:10.1103/PhysRevB.87.100401.
9. Ramasubramaniam, A.; Naveh, D. Mn-doped monolayer $MoS_2$: An atomically thin dilute magnetic semiconductor. *Phys. Rev. B - Condens. Matter Mater. Phys.* **2013**, *87*, 1–7, doi:10.1103/PhysRevB.87.195201.
10. Wu, C.-W.; Yao, D.-X. Robust p-orbital half-metallicity and high Curie-temperature in the hole-doped anisotropic $TcX_2$ (X= S, Se) nanosheets. *J. Magn. Magn. Mater.* **2019**, *478*, 68–76.
11. Coelho, P.M.; Komsa, H.; Lasek, K.; Kalappattil, V.; Karthikeyan, J.; Phan, M.; Krasheninnikov, A. V.; Batzill, M. Room-Temperature Ferromagnetism in $MoTe_2$ by Post-Growth Incorporation of Vanadium Impurities. *Adv. Electron. Mater.* **2019**, *5*, 1900044, doi:10.1002/aelm.201900044.
12. Li, B.; Xing, T.; Zhong, M.; Huang, L.; Lei, N.; Zhang, J.; Li, J.; Wei, Z. A two-dimensional Fe-doped $SnS_2$ magnetic semiconductor. *Nat. Commun.* **2017**, *8*, 1–7, doi:10.1038/s41467-017-02077-z.
13. Zhang, K.; Feng, S.; Wang, J.; Azcatl, A.; Lu, N.; Addou, R.; Wang, N.; Zhou, C.; Lerach, J.; Bojan, V.; et al. Manganese Doping of Monolayer $MoS_2$: The Substrate Is Critical. *Nano Lett.* **2015**, *15*, 6586–6591, doi:10.1021/acs.nanolett.5b02315.

14. Zhang, K.; Bersch, B.M.; Joshi, J.; Addou, R.; Cormier, C.R.; Zhang, C.; Xu, K.; Briggs, N.C.; Wang, K.; Subramanian, S.; et al. Tuning the Electronic and Photonic Properties of Monolayer MoS2 via In Situ Rhenium Substitutional Doping. *Adv. Funct. Mater.* **2018**, *28*, 1–7, doi:10.1002/adfm.201706950.
15. Fang, J.; Song, H.; Li, B.; Zhou, Z.; Yang, J.; Lin, B.; Liao, Z.; Wei, Z. Large unsaturated magnetoresistance of 2D magnetic semiconductor Fe-SnS2 homojunction. *J. Semicond.* **2022**, *43*, 92501.
16. Xing, S.; Zhou, J.; Zhang, X.; Elliott, S.; Sun, Z. Theory, properties and engineering of 2D magnetic materials. *Prog. Mater. Sci.* **2022**, 101036.
17. Li, X.; Lu, J.-T.; Zhang, J.; You, L.; Su, Y.; Tsymbal, E.Y. Spin-dependent transport in van der Waals magnetic tunnel junctions with Fe3GeTe2 electrodes. *Nano Lett.* **2019**, *19*, 5133–5139.
18. Zou, R.; Zhan, F.; Zheng, B.; Wu, X.; Fan, J.; Wang, R. Intrinsic quantum anomalous Hall phase induced by proximity in the van der Waals heterostructure germanene/Cr 2 Ge 2 Te 6. *Phys. Rev. B* **2020**, *101*, 161108.
19. Song, T.; Cai, X.; Tu, M.W.-Y.; Zhang, X.; Huang, B.; Wilson, N.P.; Seyler, K.L.; Zhu, L.; Taniguchi, T.; Watanabe, K.; et al. Giant tunneling magnetoresistance in spin-filter van der Waals heterostructures. *Science (80-. ).* **2018**, *360*, 1214–1218.
1. Burch, K.S.; Mandrus, D.; Park, J.-G. Magnetism in two-dimensional van der Waals materials. *Nature* **2018**, *563*, 47–52.
2. Gong, C.; Li, L.; Li, Z.; Ji, H.; Stern, A.; Xia, Y.; Cao, T.; Bao, W.; Wang, C.; Wang, Y.; et al. Discovery of intrinsic ferromagnetism in two-dimensional van der Waals crystals. *Nature* **2017**, *546*, 265–269, doi:10.1038/nature22060.
3. Huang, B.; Clark, G.; Navarro-Moratalla, E.; Klein, D.R.; Cheng, R.; Seyler, K.L.; Zhong, D.; Schmidgall, E.; McGuire, M.A.; Cobden, D.H.; et al. Layer-dependent ferromagnetism in a van der Waals crystal down to the monolayer limit. *Nature* **2017**, *546*, 270–273, doi:10.1038/nature22391.
4. Dietl, T.; Bonanni, A.; Ohno, H. Families of magnetic semiconductors - An overview. *J. Semicond.* **2019**, *40*, doi:10.1088/1674-4926/40/8/080301.
5. Tiwari, S.; de Put, M.L.; Sorée, B.; Vandenberghe, W.G. Magnetic order and critical temperature of substitutionally doped transition metal dichalcogenide monolayers. *npj 2D Mater. Appl.* **2021**, *5*, 1–7.
6. Fan, X.-L.; An, Y.-R.; Guo, W.-J. Ferromagnetism in transitional metal-doped MoS 2 monolayer. *Nanoscale Res. Lett.* **2016**, *11*, 1–10.
7. Zhao, X.; Chen, P.; Wang, T. Controlled electronic and magnetic properties of WSe2 monolayers by doping transition-metal atoms. *Superlattices Microstruct.* **2016**, *100*, 252–257.
8. Cheng, Y.C.; Zhu, Z.Y.; Mi, W.B.; Guo, Z.B.; Schwingenschlögl, U. Prediction of two-dimensional diluted magnetic semiconductors: Doped monolayer MoS2 systems. *Phys. Rev. B - Condens. Matter Mater. Phys.* **2013**, *87*, 2–5, doi:10.1103/PhysRevB.87.100401.
9. Ramasubramaniam, A.; Naveh, D. Mn-doped monolayer MoS2: An atomically thin dilute magnetic semiconductor. *Phys. Rev. B - Condens. Matter Mater. Phys.* **2013**, *87*, 1–7, doi:10.1103/PhysRevB.87.195201.
10. Wu, C.-W.; Yao, D.-X. Robust p-orbital half-metallicity and high Curie-temperature in the hole-doped anisotropic TcX2 (X= S, Se) nanosheets. *J. Magn. Magn. Mater.* **2019**, *478*, 68–76.
11. Coelho, P.M.; Komsa, H.; Lasek, K.; Kalappattil, V.; Karthikeyan, J.; Phan, M.; Krasheninnikov, A. V.; Batzill, M. Room-Temperature Ferromagnetism in MoTe 2 by Post-Growth Incorporation of Vanadium Impurities. *Adv. Electron. Mater.* **2019**, *5*, 1900044, doi:10.1002/aelm.201900044.
12. Li, B.; Xing, T.; Zhong, M.; Huang, L.; Lei, N.; Zhang, J.; Li, J.; Wei, Z. A two-dimensional Fe-doped SnS2 magnetic semiconductor. *Nat. Commun.* **2017**, *8*, 1–7, doi:10.1038/s41467-017-02077-z.
13. Zhang, K.; Feng, S.; Wang, J.; Azcatl, A.; Lu, N.; Addou, R.; Wang, N.; Zhou, C.; Lerach, J.; Bojan, V.; et al. Manganese Doping of Monolayer MoS2: The Substrate Is Critical. *Nano Lett.* **2015**, *15*, 6586–6591, doi:10.1021/acs.nanolett.5b02315.
14. Zhang, K.; Bersch, B.M.; Joshi, J.; Addou, R.; Cormier, C.R.; Zhang, C.; Xu, K.; Briggs, N.C.; Wang, K.; Subramanian, S.; et al. Tuning the Electronic and Photonic Properties of Monolayer MoS2 via In Situ Rhenium Substitutional Doping. *Adv. Funct. Mater.* **2018**, *28*, 1–7, doi:10.1002/adfm.201706950.
15. Fang, J.; Song, H.; Li, B.; Zhou, Z.; Yang, J.; Lin, B.; Liao, Z.; Wei, Z. Large unsaturated magnetoresistance of 2D magnetic semiconductor Fe-SnS2 homojunction. *J. Semicond.* **2022**, *43*, 92501.
16. Xing, S.; Zhou, J.; Zhang, X.; Elliott, S.; Sun, Z. Theory, properties and engineering of 2D magnetic materials. *Prog. Mater. Sci.* **2022**, 101036.
17. Li, X.; Lu, J.-T.; Zhang, J.; You, L.; Su, Y.; Tsymbal, E.Y. Spin-dependent transport in van der Waals magnetic tunnel junctions with Fe3GeTe2 electrodes. *Nano Lett.* **2019**, *19*, 5133–5139.
18. Zou, R.; Zhan, F.; Zheng, B.; Wu, X.; Fan, J.; Wang, R. Intrinsic quantum anomalous Hall phase induced by proximity in the van der Waals heterostructure germanene/Cr 2 Ge 2 Te 6. *Phys. Rev. B* **2020**, *101*, 161108.
19. Song, T.; Cai, X.; Tu, M.W.-Y.; Zhang, X.; Huang, B.; Wilson, N.P.; Seyler, K.L.; Zhu, L.; Taniguchi, T.; Watanabe,

K.; et al. Giant tunneling magnetoresistance in spin-filter van der Waals heterostructures. *Science (80-. ).* **2018**, *360*, 1214–1218.
20. Wang, J.; Sun, F.; Yang, S.; Li, Y.; Zhao, C.; Xu, M.; Zhang, Y.; Zeng, H. Robust ferromagnetism in Mn-doped MoS2 nanostructures. *Appl. Phys. Lett.* **2016**, *109*, 1–6, doi:10.1063/1.4961883.
21. Fang, Q.; Zhao, X.; Huang, Y.; Xu, K.; Min, T.; Chu, P.K.; Ma, F. Structural stability and magnetic-exchange coupling in Mn-doped monolayer/bilayer MoS 2. *Phys. Chem. Chem. Phys.* **2018**, *20*, 553–561, doi:10.1039/c7cp05988d.
22. Qi, J.; Li, X.; Chen, X.; Hu, K. Strain tuning of magnetism in Mn doped MoS2 monolayer. *J. Phys. Condens. Matter* **2014**, *26*, doi:10.1088/0953-8984/26/25/256003.
23. Li, Q.; Zhao, X.; Deng, L.; Shi, Z.; Liu, S.; Wei, Q.; Zhang, L.; Cheng, Y.; Zhang, L.; Lu, H.; et al. Enhanced Valley Zeeman Splitting in Fe-Doped Monolayer MoS2. *ACS Nano* **2020**, *14*, 4636–4645, doi:10.1021/acsnano.0c00291.
24. Briggs, N.; Subramanian, S.; Lin, Z.; Li, X.; Zhang, X.; Zhang, K.; Xiao, K.; Geohegan, D.; Wallace, R.; Chen, L.-Q.; et al. A roadmap for electronic grade 2D materials. *2D Mater.* **2019**, *6*, 22001, doi:10.1088/2053-1583/aaf836.
25. Zhou, J.; Lin, J.; Huang, X.; Zhou, Y.; Chen, Y.; Xia, J.; Wang, H.; Xie, Y.; Yu, H.; Lei, J.; et al. A library of atomically thin metal chalcogenides. *Nature* **2018**, *556*, 355–359.
26. Zhang, Y.; Yao, Y.; Sendeku, M.G.; Yin, L.; Zhan, X.; Wang, F.; Wang, Z.; He, J. Recent progress in CVD growth of 2D transition metal dichalcogenides and related heterostructures. *Adv. Mater.* **2019**, *31*, 1901694.
27. Hernandez Ruiz, K.; Wang, Z.; Ciprian, M.; Zhu, M.; Tu, R.; Zhang, L.; Luo, W.; Fan, Y.; Jiang, W. Chemical vapor deposition mediated phase engineering for 2D transition metal dichalcogenides: Strategies and applications. *Small Sci.* **2022**, *2*, 2100047.
28. Yang, L.; Wu, H.; Zhang, L.; Zhang, W.; Li, L.; Kawakami, T.; Sugawara, K.; Sato, T.; Zhang, G.; Gao, P.; et al. Highly Tunable Near-Room Temperature Ferromagnetism in Cr-Doped Layered Td-WTe2. *Adv. Funct. Mater.* 2021, *31*, doi:10.1002/adfm.202008116.
29. Fu, S.; Kang, K.; Shayan, K.; Yoshimura, A.; Dadras, S.; Wang, X.; Zhang, L.; Chen, S.; Liu, N.; Jindal, A.; et al. Enabling room temperature ferromagnetism in monolayer MoS2 via in situ iron-doping. *Nat. Commun.* **2020**, *11*, 6–13, doi:10.1038/s41467-020-15877-7.
30. Kang, K.; Fu, S.; Shayan, K.; Anthony, Y.; Dadras, S.; Yuzan, X.; Kazunori, F.; Terrones, M.; Zhang, W.; Strauf, S.; et al.  The effects of substitutional Fe-doping on magnetism in MoS 2 and WS 2 monolayers . *Nanotechnology* **2021**, *32*, 095708, doi:10.1088/1361-6528/abcd61.
31. Fu, S.; Kang, K.; Shayan, K.; Yoshimura, A.; Dadras, S.; Wang, X.; Zhang, L.; Chen, S.; Liu, N.; Jindal, A.; et al. Enabling room temperature ferromagnetism in monolayer MoS2 via in situ iron-doping. *Nat. Commun.* **2020**, *11*, 6–13, doi:10.1038/s41467-020-15877-7.
32. Yun, S.J.; Duong, D.L.; Ha, D.M.; Singh, K.; Phan, T.L.; Choi, W.; Kim, Y.M.; Lee, Y.H. Ferromagnetic Order at Room Temperature in Monolayer WSe2 Semiconductor via Vanadium Dopant. *Adv. Sci.* **2020**, *7*, 1–6, doi:10.1002/advs.201903076.
33. Shen, D.; Zhao, B.; Zhang, Z.; Zhang, H.; Yang, X.; Huang, Z.; Li, B.; Song, R.; Jin, Y.; Wu, R.; et al. Synthesis of Group VIII Magnetic Transition-Metal-Doped Monolayer MoSe2. *ACS Nano* **2022**, *16*, 10623–10631, doi:10.1021/acsnano.2c02214.
34. Tong, X.; Liu, K.; Zeng, M.; Fu, L. Vapor-phase growth of high-quality wafer-scale two-dimensional materials. *InfoMat* **2019**, *1*, 460–478.
35. Zhang, K.; Deng, D.D.; Zheng, B.; Wang, Y.; Perkins, F.K.; Briggs, N.C.; Crespi, V.H.; Robinson, J.A. Tuning transport and chemical sensitivity via niobium doping of synthetic MoS2. *Adv. Mater. Interfaces* **2020**, *7*, 2000856, doi:10.1002/admi.202000856.
36. Stolz, S.; Kozhakhmetov, A.; Dong, C.; Gröning, O.; Robinson, J.A.; Schuler, B. Layer-dependent Schottky contact at van der Waals interfaces: V-doped WSe2 on graphene. *npj 2D Mater. Appl.* **2022**, *6*, 1–5, doi:10.1038/s41699-022-00342-4.
37. Kozhakhmetov, A.; Schuler, B.; Tan, A.M.Z.; Cochrane, K.A.; Nasr, J.R.; El-Sherif, H.; Bansal, A.; Vera, A.; Bojan, V.; Redwing, J.M.; et al. Scalable Substitutional Re-Doping and its Impact on the Optical and Electronic Properties of Tungsten Diselenide. *Adv. Mater.* **2020**, *2005159*, 1–9, doi:10.1002/adma.202005159.
38. Stringfellow, G.B. *Organometallic vapor-phase epitaxy: theory and practice*; Elsevier, 1999;
39. Xu, X.; Guo, T.; Kim, H.; Hota, M.K.; Alsaadi, R.S.; Lanza, M.; Zhang, X.; Alshareef, H.N. Growth of 2D materials at the wafer scale. *Adv. Mater.* **2022**, *34*, 2108258.
40. Duan, H.; Guo, P.; Wang, C.; Tan, H.; Hu, W.; Yan, W.; Ma, C.; Cai, L.; Song, L.; Zhang, W.; et al. Beating the exclusion rule against the coexistence of robust luminescence and ferromagnetism in chalcogenide monolayers. *Nat. Commun.* **2019**, *10*, 1–9, doi:10.1038/s41467-019-09531-0.
41. Zhang, F.; Zheng, B.; Sebastian, A.; Olson, H.; Liu, M.; Fujisawa, K.; Pham, Y.T.H.; Jimenez, V.O.; Kalappattil, V.;

Miao, L.; et al. Monolayer vanadium-doped tungsten disulfide: A room-temperature dilute magnetic semiconductor. *arXiv* **2020**, 1–38.

42. Yang, L.; Wu, H.; Zhang, L.; Zhang, G.; Li, H.; Jin, W.; Zhang, W.; Chang, H. Tunable and Robust Near-Room-Temperature Intrinsic Ferromagnetism of a van der Waals Layered Cr-Doped 2H-MoTe2Semiconductor with an Out-of-Plane Anisotropy. *ACS Appl. Mater. Interfaces* **2021**, *13*, 31880–31890, doi:10.1021/acsami.1c07680.
43. Wang, Y.; Cong, C.; Yang, W.; Shang, J.; Peimyoo, N.; Chen, Y.; Kang, J.; Wang, J.; Huang, W.; Yu, T. Strain-induced direct--indirect bandgap transition and phonon modulation in monolayer WS 2. *Nano Res.* **2015**, *8*, 2562–2572.
44. Kang, W.T.; Lee, I.M.; Yun, S.J.; Song, Y. Il; Kim, K.; Kim, D.-H.; Shin, Y.S.; Lee, K.; Heo, J.; Kim, Y.-M.; et al. Direct growth of doping controlled monolayer WSe 2 by selenium-phosphorus substitution. *Nanoscale* **2018**, *10*, 11397–11402.
45. Sasaki, S.; Kobayashi, Y.; Liu, Z.; Suenaga, K.; Maniwa, Y.; Miyauchi, Y.; Miyata, Y. Growth and optical properties of Nb-doped WS $_2$monolayers. *Appl. Phys. Express* **2016**, *9*, 71201, doi:10.7567/APEX.9.071201.
46. Moody, G.; Kavir Dass, C.; Hao, K.; Chen, C.-H.; Li, L.-J.; Singh, A.; Tran, K.; Clark, G.; Xu, X.; Berghäuser, G.; et al. Intrinsic homogeneous linewidth and broadening mechanisms of excitons in monolayer transition metal dichalcogenides. *Nat. Commun.* **2015**, *6*, 8315.
47. Xiao, D.; Yao, W.; Niu, Q. Valley-contrasting physics in graphene: magnetic moment and topological transport. *Phys. Rev. Lett.* **2007**, *99*, 236809.
48. Mak, K.F.; Xiao, D.; Shan, J. Light--valley interactions in 2D semiconductors. *Nat. Photonics* **2018**, *12*, 451–460.
49. Ahmed, S.; Ding, X.; Murmu, P.P.; Bao, N.; Liu, R.; Kennedy, J.; Wang, L.; Ding, J.; Wu, T.; Vinu, A.; et al. High Coercivity and Magnetization in WSe2 by Codoping Co and Nb. *Small* **2020**, *16*, 2–9, doi:10.1002/smll.201903173.
50. Kabiraj, A.; Kumar, M.; Mahapatra, S. High-throughput discovery of high Curie point two-dimensional ferromagnetic materials. *npj Comput. Mater.* **2020**, *6*, 35.
51. Yun, S.J.; Duong, D.L.; Ha, D.M.; Singh, K.; Phan, T.L.; Choi, W.; Kim, Y.-M.; Lee, Y.H. Ferromagnetic order at room temperature in monolayer WSe2 semiconductor via vanadium dopant. *Adv. Sci.* **2020**, *7*, 1903076.
52. Room-Temperature Ferromagnetism in MoTe2 by Post-Growth Incorporation of Vanadium Impurities.pdf.
53. Ahmed, S.; Ding, X.; Bao, N.; Bian, P.; Zheng, R.; Wang, Y.; Murmu, P.P.; Kennedy, J.V.; Liu, R.; Fan, H.; et al. Inducing High Coercivity in MoS2 Nanosheets by Transition Element Doping. *Chem. Mater.* **2017**, *29*, 9066–9074, doi:10.1021/acs.chemmater.7b02593.
54. Advanced Science - 2021 - Lin - Controllable Thin-Film Approaches for Doping and Alloying Transition Metal Dichalcogenides.pdf.